\documentclass{article}

\usepackage{arxiv}

\usepackage[utf8]{inputenc} % allow utf-8 input
\usepackage[T1]{fontenc}    % use 8-bit T1 fonts
\usepackage{hyperref}       % hyperlinks
\usepackage{url}            % simple URL typesetting
\usepackage{booktabs}       % professional-quality tables
\usepackage{amsfonts}       % blackboard math symbols
\usepackage{nicefrac}       % compact symbols for 1/2, etc.
\usepackage{microtype}      % microtypography
\usepackage{lipsum}		% Can be removed after putting your text content
\usepackage{graphicx}
\usepackage{doi}
\usepackage{biblatex}
\title{SeqMate: A Novel Large Language Model Pipeline for Automating RNA Sequencing}

\author{ \href{https://orcid.org/0009-0005-8025-0724}{\includegraphics[scale=0.06]{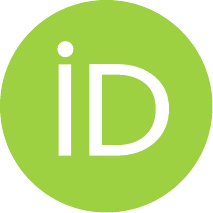}\hspace{1mm}Devam Mondal}\\
	Center for Complex Systems and Enterprises\\
	Stevens Institute of Technology\\
	Hoboken, NJ \\
	\texttt{dmondal@stevens.edu} \\
	\And
	{\hspace{1mm}Atharva Inamdar} \\
	\texttt{atharva.inamdar006@gmail.com} \\
}

\renewcommand{\headeright}{Tech. Report}
\renewcommand{\undertitle}{Technical Report}
\hypersetup{
pdftitle={A template for the arxiv style},
pdfsubject={q-bio.NC, q-bio.QM},
pdfauthor={David S.~Hippocampus, Elias D.~Striatum},
pdfkeywords={First keyword, Second keyword, More},
}

\begin{document}
\maketitle

\begin{abstract}

	RNA sequencing techniques, like bulk RNA-seq and Single Cell (sc) RNA-seq, are critical tools for the biologist looking to analyze the genetic activity/transcriptome of a tissue or cell during an experimental procedure. Platforms like Illumina's next-generation sequencing (NGS) are used to produce the raw data for this experimental procedure. This raw FASTQ data must then be prepared via a complex series of data manipulations by bioinformaticians. This process currently takes place on an unwieldy textual user interface like a terminal/command line that requires the user to install and import multiple program packages, preventing the untrained biologist from initiating data analysis. Open-source platforms like Galaxy have produced a more user-friendly pipeline, yet the visual interface remains cluttered and highly technical, remaining uninviting for the natural scientist. To address this, SeqMate is a user-friendly tool that allows for one-click analytics by utilizing the power of a large language model (LLM) to automate both data preparation and analysis (differential expression, trajectory analysis, etc). Furthermore, by utilizing the power of generative AI, SeqMate is also capable of analyzing such findings and producing written reports of upregulated/downregulated/user-prompted genes with sources cited from known repositories like PubMed, PDB, and Uniprot. 

\end{abstract}

\section{Introduction}

	Processing RNA-seq data is tedious. Converting the FASTQ file output of bulk-seq or chip-seq involves numerous file conversions and data processing involving numerous packages in a process that remains unautomated and inaccessible to the biologist untrained in bioinformatics. Many advances in process automation, especially utilizing the technological boom in generative AI and large language models, have not been implemented in the bioinformatics space. SeqMate is looking to change that by utilizing an accountable (cites sources and actively prompted against hallucinations) LLM that automates bioinformatic analysis like trajectory analysis, differential expression, and more. Currently, the tool is designed and tested for differential expression starting with raw FASTQ data, but SeqMate is envisioned as a suite of multiple bioinformatic processes that provides a one-click option for the untrained biologist to spearhead data analysis.

\section{RNA Sequencing: A Brief Biological Context}

RNA sequencing involves high-throughput sequencing of RNA nucleic acid sequences and amounts. RNA-seq has led to many scientific advances, such as identifying regulatory regions, biomarkers, and mutations \cite{Deshpande2023}. Within the family of RNA sequencing exist two distinct processes, bulk RNA sequencing and single-cell RNA sequencing. Bulk RNA sequencing allows a researcher to look at the transcriptome of a sample/tissue, commonly by using NGS technology like the Illumina sequencing platform. Single-cell seq differs in that instead of looking at the transcriptome of an entire tissue, one can analyze the transcriptome of an individual cell by using microfluidics to separate individual cells and barcoding the RNA of each cell with unique identifiers, allowing researchers to identify the genetic activity of various cell sub-populations within a sample \cite{Li2021}. Both these technologies allow researchers to take snapshots of the genetic activity of a tissue/cell and are powerful tools to determine changes in expression levels of genes after any experimental treatment. Coupling multiple scRNA-seqs/bulk-seqs over time allows a powerful time series analysis of the expression levels of an experimental sample \cite{Deshpande2023}. 

\section{Role of Bioinformatics in RNA Sequencing}

RNA-seq is a powerful high throughput sequencing tool that can produce large amounts of data, yet there is still a large amount of tedious data processing that remains before meaningful insights can be gleaned. To delineate, the traditional bioinformatics process from raw RNA-seq data looks like the following: 

\begin{enumerate}
    \item Perform quality control.
    \item Cut adapter regions and align with reference genome.
    \item Perform file conversion and produce count matrices. 
    \item Normalize counts.
    \item Perform differential expression analysis.
    \item Relate upregulation/downregulation findings with relevant biological pathways.
    \item Create write-ups/visuals.
\end{enumerate}

After such analysis, one can meaningfully interpret and extrapolate the results of RNA-seq to their experimental design \cite{Wu2017}.

\section{Current Limitations of Bioinformatics Suites}

Currently, bioinformatics analysis involves knowledgeable professionals running various open-source packages on some form of textual user interface (command line/terminal). Conducting such analysis involves importing/downloading the correct packages, uploading the data, running the correct commands, and downloading the processed data to the right location for every step of a multi-step process that takes place in a lay-person unfriendly environment. This cryptic process obfuscates natural scientists like cellular biologists from conducting their own analysis and bottlenecks the scientific process. Open-source platform Galaxy has made a bioinformatics analysis pipeline, but its usability is limited by a confusing visual interface. While Galaxy has created a powerful tool that streamlines bioinformatics for the trained specialist, it is still too complex and technical to be quickly intuitive for the biologist. Additionally, the pipelines Galaxy provides are generic and do not cater to specific use cases, instead serving as general guides \cite{Afgan2018}.

\section{Our Approach}

In our goal to automate the process of RNA Sequencing, we aimed to create an autonomous system that was capable of receiving user input, running the appropriate intermediate steps (generating genome indices, creating appropriate metadata for differential expression analysis, filtering genome indices, etc.), and providing an output. For this particular use case, we decided that the user input would be a series of FASTQ files, the appropriate genome, thresholds for log fold change and p-value, specification of control and experimental FASTQ files, and the output would be a series of genes demonstrating greatest differential expression based on user-defined thresholds (log fold change and p-value) and detailed information about them. 

Such a system would require a "logical driver" that conducts decision-making (picking the appropriate files either from the user or from generated files in other steps of the intermediate process, running the appropriate code/terminal commands in an environment, generating graphs, providing narration, etc.). We chose to use a large language model to be this component of our system due to their ability to understand data through insight generation. 

However, vanilla LLMs have two main challenges. First, they are worse at solving complex reasoning problems that require decision making at intermediate steps compared to simple, single-step tasks \cite{mondorf2024accuracyevaluatingreasoningbehavior}. Additionally, they lack additional functionalities necessary within this system, such as the ability to run open-source libraries on native environments as well as search online databases/browsers (such as the NCBI GenBank Database and Ensembl). Therefore, we couple the large language model with a variety of external tools that enable the aforementioned tasks, thus creating an \textbf{LLM agent}. Furthermore, we make this agent using the open-source package LangChain. LangChain enables us to improve an LLM's ability to solve complex, reasoning problems by forcing the LLM to use chain-of-thought reasoning (where it assesses the given prompt/query, breaks it down into intermediate steps that are rationale, and executes them using a variant of the ReAct paradigm) \cite{huang-chang-2023-towards} \cite{wei2023chainofthoughtpromptingelicitsreasoning} \cite{yao2023reactsynergizingreasoningacting}. Figure 1 demonstrates this coupling.

\begin{figure}[hbt!]
    \centering
    \includegraphics[scale=0.35]{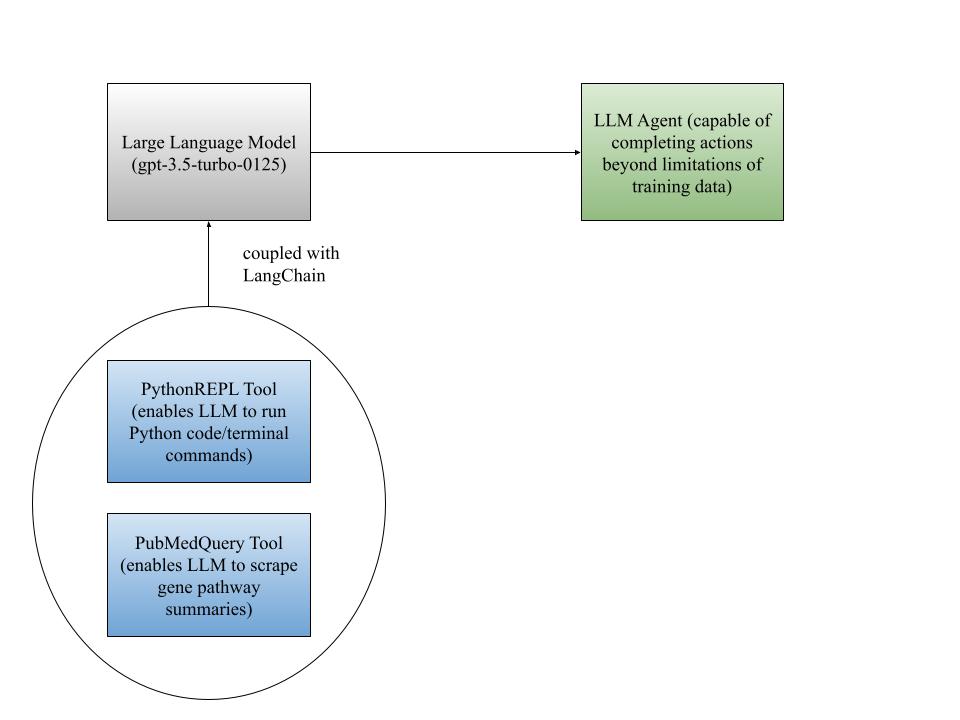}
    \caption{Visualization of LLM agent creation.}
    \label{fig:enter-label}
\end{figure}

With this agent, a combination of an LLM and external tools, we automate each intermediate step through prompt engineering that outlines what files it must use, what package is appropriate, and what files it must output. We also introduce few-shot and one-shot prompting with code snippets in order to increase the success rate of each prompt evocation. Additionally, the LLM's decision-making abilities increase when integrated with LangChain by placing emphasis on chain-of-thought reasoning when approaching each prompt. With these prompts, the agent is able to follow the following pipeline without user interaction:

\begin{enumerate}
    \item Open the user-provided FASTQ files.
    \item Remove low-quality regions and adapters using \texttt{cutadapt}.
    \item Provide quality control statistics for edited FASTQ files using \texttt{bio}, generating graphics and providing narration.
    \item Generate a genome index for a user-provided genome using \texttt{hisat}.
    \item Align edited FASTQ files to the aforementioned genome index using \texttt{hisat}, producing a SAM output file.
    \item Convert the SAM output file to a BAM file using \texttt{pysam}.
    \item Download the genome annotation file for the genome.
    \item Create a count matrix using the BAM files and the genome annotation file using \texttt{featureCounts}.
    \item Edit the counts table using appropriate matrix operations.
    \item Create metadata needed for differential expression analysis
    \item Run differential expression analysis using the count matrix and metadata using \texttt{pydeseq2}.
    \item Filter outlier genes based on user thresholds mentioned earlier.
    \item Generate insights on outlier genes using external databases and provide a detailed, cohesive summary with citations.
    
\end{enumerate}
    
It is important to note that the agent utilizes the source machine's environment in order to run the open-source packages. It is thus able to utilize packages present in \texttt{pip} and \texttt{conda} environments that are set up on the source machine, make alterations through installations, and switch to other environments.

\section{Limitations and Future Work}

Due to intrinsic noise that exists in training data, LLMs are susceptible to hallucinations where they produce factually incorrect results. Though our prompt engineering limits this through one-shot/zero-shot prompts, the pipeline sometimes generates inaccurate results. We hope to generate statistics for hallucinations (frequency, type, etc.) in a follow-up report.

Additionally, generating a genome index through \texttt{hisat} is currently computationally expensive. However, this should not be an issue when the pipeline is run on a desktop or server.

Moreover, for the LLM in the system, we currently rely on OpenAI's \texttt{gpt-3.5-turbo-0125} model. This may pose a privacy concern, as aspects of the data supplied to the agent are sent to OpenAI through an API. In the future, we would like to experiment with open-source LLMs that can be ran locally.

It is also important to create a graphical user interface for users to interact with the LLM agent, and we also plan on extrapolating our pipeline to other bioinformatics processes. 

\section{Conclusion}

In summary, SeqMate is an automated pipeline that simplifies RNA-seq data processing and analysis built for the natural scientist. Unlike traditional "pipelines" existing bioinformatics suites provide, SeqMate is custom to every use-case by leveraging a large language model agent. In the future, SeqMate looks to expand the pipeline to include other analytical methods following the same user-friendly design policy.

\printbibliography

\end{document}